\documentclass[]{aa}
\usepackage{graphicx}
\usepackage{txfonts}
\usepackage{natbib,twoopt}
\bibpunct{(}{)}{;}{a}{}{,}
\usepackage[breaklinks=true]{hyperref} 
\bibpunct{(}{)}{;}{a}{}{,} 
\makeatletter
\newcommandtwoopt{\citeads}[3][][]{\href{http://adsabs.harvard.edu/abs/#3}%
{\def\hyper@linkstart##1##2{}%
\let\hyper@linkend\@empty\citealp[#1][#2]{#3}}}
\newcommandtwoopt{\citepads}[3][][]{\href{http://adsabs.harvard.edu/abs/#3}%
{\def\hyper@linkstart##1##2{}%
\let\hyper@linkend\@empty\citep[#1][#2]{#3}}}
\newcommandtwoopt{\citetads}[3][][]{\href{http://adsabs.harvard.edu/abs/#3}%
{\def\hyper@linkstart##1##2{}%
\let\hyper@linkend\@empty\citet[#1][#2]{#3}}}
\newcommandtwoopt{\citeyearads}[3][][]%
{\href{http://adsabs.harvard.edu/abs/#3}
{\def\hyper@linkstart##1##2{}%
\let\hyper@linkend\@empty\citeyear[#1][#2]{#3}}}
\makeatother

\begin{document}


\title{Astrometric observations of Phobos and Deimos during the 1971 opposition of Mars}
\author{V. Robert\inst{1}\fnmsep\inst{2}\and V. Lainey\inst{1}\and D. Pascu\inst{3}\fnmsep\thanks{USNO retired}\and J.-E. Arlot\inst{1}\and J.-P. De Cuyper\inst{4}\and V. Dehant\inst{4}\and W. Thuillot\inst{1}}
\institute{
Institut de M\'ecanique C\'eleste et de Calcul des \'Eph\'em\'erides IMCCE, Observatoire de Paris, UMR 8028 du CNRS, UPMC, Univ. Lille 1, 77 avenue Denfert-Rochereau, 75014 Paris, France
\and
Institut Polytechnique des Sciences Avanc\'ees IPSA, 7-9 rue Maurice Grandcoing, 94200 Ivry-sur-Seine, France
\and
United States Naval Observatory USNO, 3458 Massachusetts Ave NW, Washington, DC 20392, USA
\and
Royal Observatory of Belgium ROB, avenue Circulaire 3, B-1180 Uccle, Belgique
}
\date{Received 12 June 2014 / Accepted 11 September 2014.}

 
\abstract
{Accurate positional measurements of planets and satellites are used to improve our knowledge of their dynamics and to infer the accuracy of planet and satellite ephemerides.}
{In the framework of the FP7 ESPaCE project, we provide the positions of Mars, Phobos, and Deimos taken with the U.S. Naval Observatory 26-inch refractor during the 1971 opposition of the planet.}
{These plates were measured with the digitizer of the Royal Observatory of Belgium and reduced through an optimal process that includes image, instrumental, and spherical corrections to provide the most accurate data.}
{We compared the observed positions of the planet Mars and its satellites with the theoretical positions from INPOP10 and DE430 planetary ephemerides, and from NOE and MAR097 satellite ephemerides. The rms residuals in RA and DEC of one position is less than 60 mas, or about 20 km at Mars. This accuracy is comparable to the most recent CCD observations. Moreover, it shows that astrometric data derived from photographic plates can compete with those of old spacecraft (Mariner 9, Viking 1 and 2).}{}
\titlerunning{Astrometric observations of Phobos and Deimos}
\authorrunning{V. Robert et al.}
\maketitle

 
\section{Introduction}
The European Satellites Partnership for Computing Ephemerides ESPaCE project aims at strengthening the collaboration and at developing new knowledge, new technology, and products for the scientific community in the domains of the development of ephemerides and reference systems for natural satellites and spacecraft. Part of the ESPaCE program aims at providing accurate positional measurements of planets and satellites that will be used to compute new orbital ephemerides. Since we had demonstrated that a precise digitization and a new astrometric reduction of old photographic plates could provide very accurate positions \citepads{2011MNRAS.415..701R}, the project leaders chose to consider such observations as a significant task for the ESPaCE program.

At the end of the 1960s a satellite-observing program was initiated at the United States Naval Observatory (USNO, Washington DC, USA) to obtain high-precision observations of the Martian system. The observations employed a long-focus instrument to provide astrometric stability and the option of a filter mask on Mars that allowed simultaneous measurement of the primary and the faint satelites \citepads{1975Icar...25..479P}. Photographic plates were taken during the 1971 opposition of Mars with the USNO 26-inch refractor by \citetads{1979nasm.conf...17P}. Some of these plates were measured manually and the resulting positions used in support of Martian space missions, as well as to study the dynamics of the satellites \citepads{2007A&A...465.1075L}. These observations were limited to the positions of the satellites relative to Mars rather than equatorial (RA, DEC) positions since dense, accurate star catalogs were not yet available. Thus our intention was to use the methods developed in the USNO Galilean plates analysis in \citet{robertphd} and \citetads{2011MNRAS.415..701R} to analyze the set of the USNO Martian observations. It is noteworthy that the 1971 epoch corresponds to that of the Mariner 9 mission and to the beginning of the Viking program, thus permitting a comparison with their positional accuracies.

 
\section{Observations}

\subsection{The astrophotographic plates}

Forty years ago, the satellite motions were still not well-known. In particular, accelerations in their longitudes needed to be investigated, and it also appeared necessary to obtain new astrometric positions of the satellites because of the needs for new space missions under preparation.

Getting accurate positions of Phobos and Deimos from ground-based observations has always been difficult. Their apparent magnitude of 11.3 and 12.4, respectively, are quite faint compared to Mars, about -2.0. Furthermore, their orbits are very close to the planet, and the scattered light from Mars increases the level of the sky background and affects the satellite positions, systematically shifting them toward the planet's center. To minimize these effects, suitable reduction techniques had to be used.

Photographic plates used by Dan Pascu were Kodak 103aJ glass plates. The observations set was taken within five nights from August 7, 1971 to September 3, 1971 \citepads{1979nasm.conf...17P}. Each plate contains two exposures shifted in the RA direction, and eighty plates resulting in 159 exposures were taken. The exposure time of the photographic plates are 10-60 seconds. A small, neutrally transmitting Nichrome film filter was deposited on a Schott GG14 yellow filter. The Nichrome filter was placed over the planet in the observations. This technique allows the primary to be measured without affecting its resolution, and it was used in combination with a V filter to fit in the right wavelengths for a refracting telescope. The field of view is 57 arcmin on the x-axis and 43 arcmin on the y-axis, with a mean angular scale of 20.851 arcsec/mm at full aperture.

\begin{figure}
	\centering
	\includegraphics[width=\columnwidth]{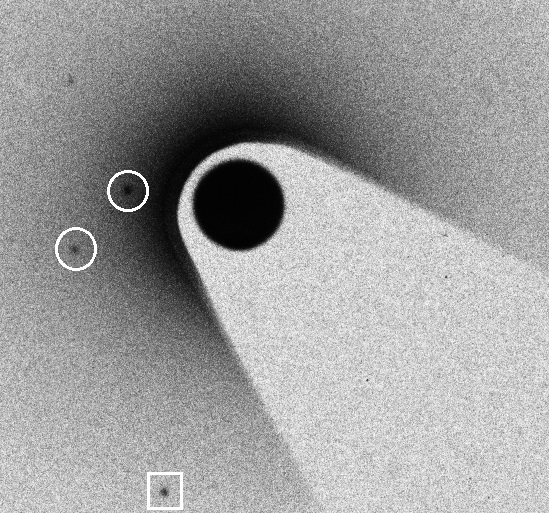}
	\caption{Center of the digitization (negative) of the USNO Martian plate n$^\circ$03017. Phobos and Deimos are displayed in the white circles. Phobos is closest to the planet. A star is displayed in the white square at the bottom of the image.}
	\label{fig001}
\end{figure}

Forty-nine plates resulting in 95 exposures were selected and transmitted to ROB to be digitized \citepads{2011MNRAS.415..701R,2011ASPC..442..301D}. We were able to make accurate measurements for 87 positions of Mars and Phobos and 58 positions of Deimos. This last number is different because the planetary filter obscured the image of the satellite during the fourth night of observations. As an example, Figure~\ref{fig001} shows the center of the digitized (negative) USNO Martian plate n$^\circ$03017 of August 13, 1971, which is a typical digitized image. Phobos and Deimos are displayed there.

\subsection{Plate measurement}

The measured (x,y) coordinates of Mars, its satellites, and the reference stars were extracted from the mosaic FITS images by a suitable process. First, we used the Source Extractor software \citepads{1996A&AS..117..393B} to create a list of objects that were detected on the images. The stars assumed to be present in the field were identified from the chosen catalogs. On the other hand, we developed special IDL software to extract planet and satellite positions. Because the observations were taken at the opposition and minimum distance from Earth, the Mars positions were determined with a two-dimensional limb fitting by least squares. We used the method described in \citeads{2013A&A...551A.129T} to determine limb points and fit the shape of the planet with a configurable ellipsoid. The satellite positions were determined with a two-dimensional Gaussian profile fitting once the sky background around each of the two was subtracted with a 3D surface fitting by least squares. Even if the planet was filtered to reduce its brightness, a high non-regular sky background gradient in the satellites' area still had to be removed (see Figure~\ref{fig001}) to determine the most precise positions. The obtained plate positions were finally corrected for the optical distortion introduced by the ROB digitizer optical unit during the digitization \citepads{2011MNRAS.415..701R}.

\subsection{Astrometric calibration}

The observations contain two to nine UCAC4 reference stars \citepads{2013AJ....145...44Z}, and the astrometric reductions were performed using suitable four constant functional models \citep{robertphd} to provide equatorial (RA, DEC) astrometric positions of the planet and its satellites. In fact, because simultaneously made cluster Praesepe plates were available for this set, we demonstrated that it was possible to reduce the observations with a unique scale factor $\rho$ for the isotropic scale and a unique orientation $\theta$. We also were able to estimate the contribution of instrumental effects. That is why mean scale $\rho$, mean orientation $\theta$, and offsets $\Delta_x$ and $\Delta_y$ were modeled. Estimated coma-magnitude terms $C_x$ and $C_y$, object magnitude $m$, average object magnitudes $m_0$, and temperature contribution terms $\epsilon_1$, $\epsilon_2$, and $\epsilon_3$ were also introduced in the determination of the tangential $(X,Y)$ coordinates for each satellite and planet:
\begin{equation}
	\left\lbrace
	\begin{array}{ll}
	X= & \rho\cos\theta\cdot x-\left(\rho+\epsilon_1\sin\left(\epsilon_2t+\epsilon_3\right)\right)\sin\theta\cdot y+\Delta_x+\\
	& C_x\cdot x\cdot \left(m-m_0\right)\\
	Y= & \rho\sin\theta\cdot x+\left(\rho+\epsilon_1\sin\left(\epsilon_2t+\epsilon_3\right)\right)\cos\theta\cdot  y+\Delta_y+\\
	& C_y\cdot y\cdot \left(m-m_0\right).
	\end{array}
	\right.
\label{eq1}
\end{equation}

Depending on the nights of observation, the observed sky fields and thus the local star catalog densities were different. As a  consequence, the number of UCAC4 reference stars could be too limited to allow for a good calibration of the images. When not enough UCAC4 referenced stars were available, we chose to combine the UCAC4 catalog with the NOMAD catalog \citepads{2004AAS...205.4815Z} in order to provide the maximum number of (RA, DEC) astrometric positions as possible.

Before adjustment, the measured coordinates of the objects were corrected for spherical effects, such as the aberration of light and the light deflection \citepads{1989AJ.....97.1197K}, the phase effect according to Lindegren's theory \citepads{1977A&A....57...55L}, and the total atmospheric refraction \citep{robertphd}. All our observations are equatorial (RA, DEC) astrometric positions obtained from tangential $(X,Y)$ coordinates by using the gnomonic inverse projection, and determined in an ICRS geocentric reference frame to be easily compared with the most recent ephemerides.


\section{Positioning results}

We compared the observed positions of Mars, Phobos, and Deimos with their theoretical computed positions given by the INPOP10 planetary ephemeris \citepads{2011CeMDA.111..363F} and NOE satellite ephemerides \citepads{2007A&A...465.1075L}. We then focused on individual observations for which the (O-C)s of the planet and the satellites were independently less than the $3\sigma$ value of their rms (O-C) in right ascension and declination. This concerns 73 positions of Mars, 71 positions of Phobos, and 51 positions of Deimos. In the list available in electronic form at the CDS\footnote{\object{Full Table~\ref{tab001} is only available in electronic form at the CDS via anonymous ftp to \url{cdsarc.u-strasbg.fr} (\url{130.79.128.5}) or via \url{http://cdsweb.u-strasbg.fr/cgi-bin/qcat?J/A+A/572/A104}}}, the corresponding geocentric observed positions refer to the ICRF, and the mean time of observation is given in Barycentric Dynamical Time TDB. Starting from the left hand column we provide the object name, the mean TDB date of observation in Julian Days, the geocentric observed right ascension, and declination in degrees. Table~\ref{tab001} gives an extract of this list.

\begin{table}
	\caption{Extract from the astrometric positions list of Mars, Phobos, and Deimos available in electronic form at the CDS.}
	\label{tab001}      
	\centering          
	\begin{tabular}{c c c c}
		\hline\hline       
		Object & Date (TDB) & RA (deg.) & DEC (deg.)\\ 
		\hline
		Mars & 2441170.676697 & 322.898652 & -21.835786\\
		Phobos & 2441170.676697 & 322.908902 & -21.834846\\
		Deimos & 2441170.676697 & 322.923583 & -21.832599\\
 		\hline
	\end{tabular}
\end{table}

\begin{figure}
	\centering
	\includegraphics[width=\columnwidth]{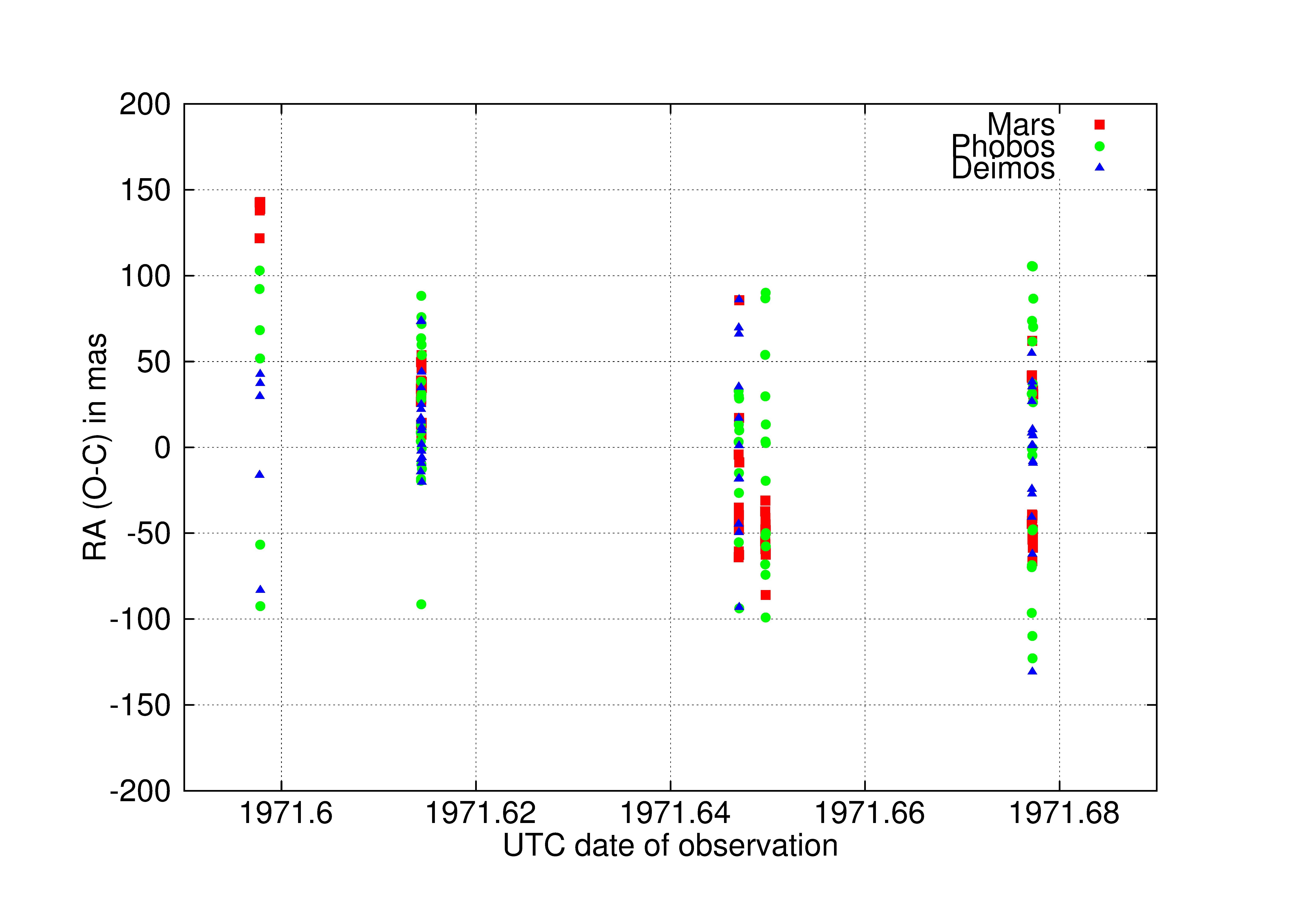}
	\includegraphics[width=\columnwidth]{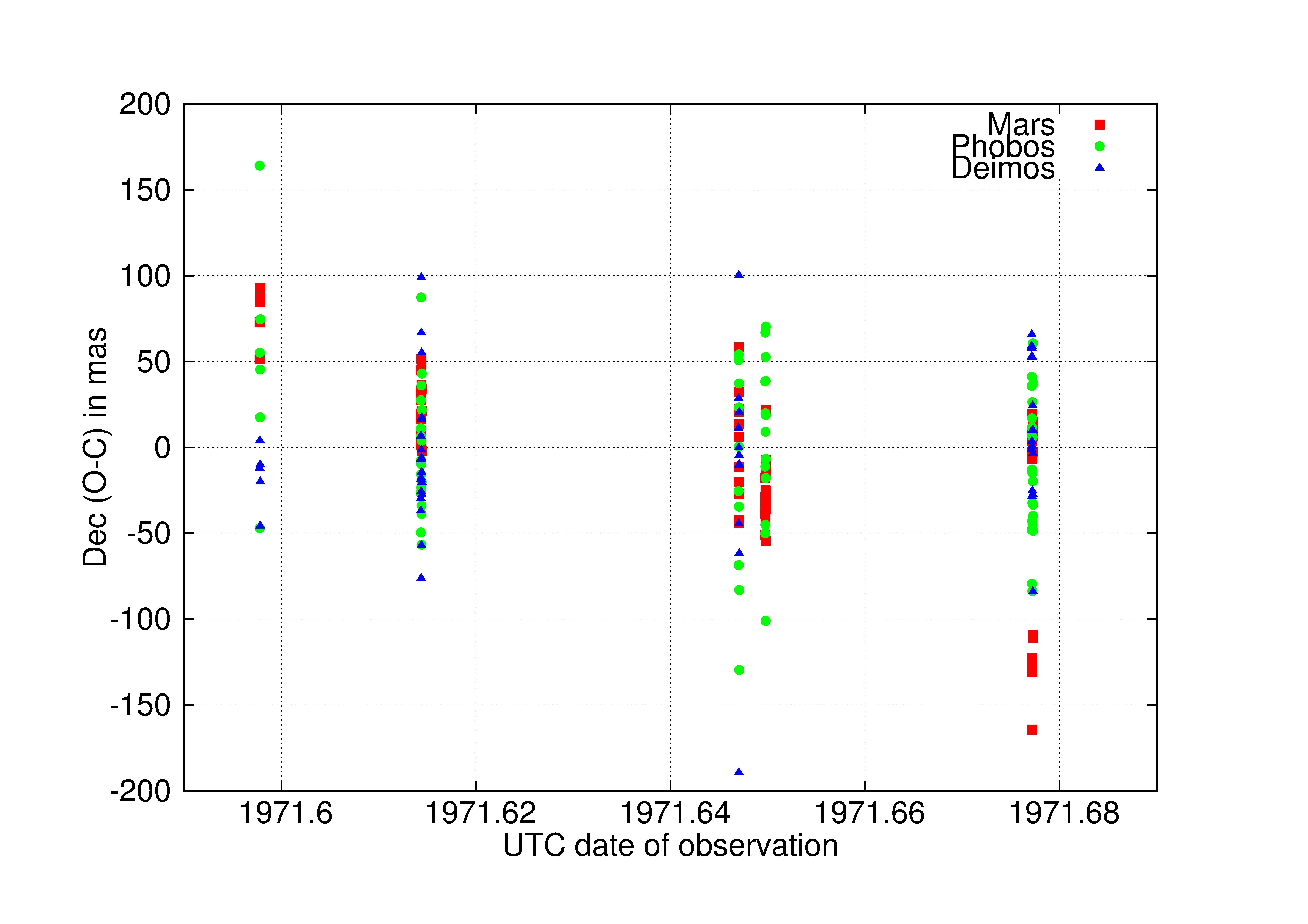}
	\caption{(RA, DEC) (O-C) according to the NOE and INPOP10 ephemerides. The $X$ axis shows the UTC date of observation and $Y$ axis the RA and DEC (O-C)s. Red  squares denote Mars, green circles Phobos, and blue triangles Deimos.}
	\label{fig002}
\end{figure}

The (O-C) distributions of residuals in equatorial right ascension and declination are provided in Figure~\ref{fig002} and Table~\ref{tab002}. They show the difference of (RA, DEC) coordinates for individual planet and satellites, hence the observed positions versus positions calculated from the INPOP10 ephemeris. Offsets for each night are very small, except for the second night of observation in the RA coordinate. Local systematic errors of the reference star catalogs could explain this bias. In fact, the epoch difference between these 1971 plates and the central epoch of the reference stars is up to about 20 years. Expected systematic errors of the reference stars at a 1971 epoch are about 10-40 mas. The data shown in Figure~\ref{fig002} span about a month, that is to say, 15$\degr$ along the path of Mars. Thus the sets of reference stars are different. With about two to nine reference stars per field, the astrometric reduction may be affected by accidental errors of individual reference stars. We assume that the exposure timing can be ruled out because the dome clock was calibrated each night to the USNO masterclock, and the camera shutter could be opened and closed on integral seconds with a precision of 0.2-0.3 seconds. The time of mid-exposure thus has a precision better than 0.5 seconds.

\begin{table*}
	\caption{Details of the (RA, DEC) (O-C) in mas according to the NOE/INPOP10 ephemerides, and to the MAR097/INPOP10 ephemerides.}
	\label{tab002}      
	\centering          
	\begin{tabular}{c c c c c || c c c c}
		\hline\hline       
		& $\overline{(O-C)_{\alpha\cos\delta}}$ & $\sigma_{\alpha\cos\delta}$ & $\overline{(O-C)_{\delta}}$ & $\sigma_{\delta}$ & $\overline{(O-C)_{\alpha\cos\delta}}$ & $\sigma_{\alpha\cos\delta}$ & $\overline{(O-C)_{\delta}}$ & $\sigma_{\delta}$\\ 
		& NOE & NOE & NOE & NOE & MAR097 & MAR097 & MAR097 & MAR097\\
		\hline
		Mars & -1.8 & 56.9 & -5.3 & 50.7 & -1.8 & 56.9 & -5.3 & 50.7\\
		Phobos & 6.0 & 58.3 & -1.7 & 48.6 & 6.0 & 58.4 & 1.2 & 48.9\\
		Deimos & 4.3 & 41.9 & -3.1 & 47.3 & -0.2 & 41.3 & -8.2 & 47.0\\
 		\hline
	\end{tabular}
\end{table*}

The average (O-C) values for the observations made during the 1971 opposition of Mars are very low in RA and DEC coordinates. The external error of INPOP10 for Mars is supposed to be less than 1 mas \citepads{2011CeMDA.111..363F}, and we may deduce from our results that this external error is realistic for both coordinates. However, the order of magnitude of the offsets are remarkably low. Such accuracy was only reached once with photographic plates in the analysis of the USNO Galilean observations \citep{robertphd}. To estimate the influence of the planetary ephemeris on the results, we computed the difference of observed positions versus positions calculated from the DE430 ephemeris \citepads{2014IPNPR.196C...1F}. Results are quite similar to those calculated from the INPOP10 ephemeris, so we may deduce that both dynamical models are equally well-constrained for this period and system. The external error of DE430 for Mars is also supposed to be less than 1 mas, which is in good agreement with our data. We finally compared the observed positions of Phobos and Deimos with their theoretical positions given by the most recent MAR097 JPL ephemerides \citepads{2010AJ....139..668J}. The average (O-C) differences in RA and DEC coordinates are below 5 mas. The respective rms (O-C) differences are below 0.5 mas. It is interesting to see that these observations are very useful for evaluating the accuracy of the ephemerides of Mars and its satellites since they have not been used for the fit of the numerical integrations leading to INPOP10 and DE430, and to NOE and MAR097, for example.

In addition, we can see that equatorial $\sigma_{\alpha\cos\delta}$ dispersion values are higher than equatorial $\sigma_\delta$, which is in good agreement with the fact that the apparent motion of the system is essentially on right ascension, but not for Deimos. This error is not as systematic as we could find from USNO Jupiter observations \citepads{2011MNRAS.415..701R} or USNO Saturn observations \citepads{1990AJ.....99.1974P}, and no causes were found in the photographic observations or in the reduction method. \citetads{1992A&AS...96..485C} found similar errors in his sets, but the question remained. The observations used to fit Phobos and Deimos models of motion could explain this error because the satellite models were mainly constrained by the Mariner 9 \citepads{1989A&A...216..284D} and Viking 1, 2 \citepads{1988A&A...201..169D} spaceraft data before the 80s. But their measurement accuracies were limited because of uncertainties in the spacecraft positions. Moreover, the model of motion of Phobos was better constrained both in right ascension and declination coordinates afterward, thanks to the Phobos 2 mission \citepads{1991A&A...244..236K}. In comparison, the Deimos' observations with Phobos 2 were rather poor. This could explain why the effect is only visible with Deimos data for the 1971 period, so further investigations are planned.

The key point is that the rms (O-C) for all observations of Mars is 54.0 mas, for Phobos, 53.6 mas, and for Deimos, 44.8 mas. These average rms (O-C) correspond to our observation accuracies at the opposition of 1971 and can be compared to those of previous Earth-based observations. \citetads{1992A&AS...96..485C} made CCD observations during the 1988 opposition with the one-meter telescope at Pic-du-Midi. Because he did not use a suitable filtering to reduce the brightness of the planet, he was not able to measure the center of Mars so that all his observations were intersatellite positions. Colas found an overall rms (O-C) of 60 mas, while our overall intersatellite rms (O-C) is 34 mas. Earlier, \citetads{1975Icar...25..479P} made his first photographic observations of Mars at the opposition of 1967 with the USNO 61-inch reflector. The observation technique was similar to the one presented in this paper. Pascu used astrophotographic field plates to calibrate and reduce his observations and then provided spherical (RA, DEC) astrometric positions. He found a mean error of 80 mas for Phobos and 100 mas for Deimos.

Several factors contributed to the increase in precision and the most important was undoubtedly the digitization of the photographic plates. Thus the background halo, in which Phobos was embedded, could be removed and model centroids applied to the images. The Nichrome film filter made it possible to measure the position of the planet's image but probably made little contribution to the precision in the positions of the satellites. Another factor was the accurate and dense star catalog used, which made it possible to obtain precision RA and DEC While the star catalog also contributed to the calibration of the plates, that contribution was more important for the large-scale system of Jupiter and Saturn \citep{robertphd}.

In the present work, we found an overall rms (O-C) of about 50 mas, that is to say 17 km. This could finally be compared to spacecraft accuracies of the same epoch. In fact, the 1971 epoch corresponds to that of the last Mariner 9 Martian mission and to the beginning of the Viking program. Mariner 9 went into orbit about Mars in November 1971. The accuracy of its data set has been first estimated to be 10 km \citepads{1989A&A...216..284D}. Viking 1 and 2 went into orbit about Mars in June and August 1976. The accuracy of their data set has been first estimated to be 4 km \citepads{1989A&A...216..284D}. Nevertheless, such numbers do not consider the error on the spacecraft position itself. As a result, the rms (O-C) of these spacecraft measurements range between 4.5 km and 25 km \citepads{2007A&A...465.1075L}. Moreover, recent observations of Deimos by Phobos 2 have a similar error of 14-16 km. Since spacecraft observations are used to acquire several images on the same orbit, errors on their position induce non-Gaussian noise due to the varying distance to the imaged moon and the sometimes limited number of observations. As a consequence, we conclude that astrometric data from photographic plates with a new treatment can provide observations of Mars moons similar in accuracy to old space data but with a much more Gaussian error.


\section{Conclusion}

We analyzed a full series of photographic plates of the Martian satellites taken at USNO during the 1971 opposition of Mars and extracted all the important information contained in the plate data. We were able to correct for instrumental and spherical effects during the reduction, thereby decreasing the number of unknown parameters. Thanks to the new technologies and new astrometric catalogs, we were able to provide astrometric positions with never achieved accuracy for Mars moons in the 1970s from the ground. We obtained equatorial RA and DEC positions of Phobos and Deimos, allowing us to deduce positions of Mars indirectly through accurate ephemerides of the satellites, but we were also able to obtain positions of the planet directly. We obtained residuals between a few tens of mas, to better than 100 mas. This provides a positioning accuracy that is very close to that of spacecrafts from the same epoch but a more numerous spread on a longer interval of time with much more Gaussian error.

The USNO archive contains more plates taken in the interval 1967-1998 and the present work has demonstrated that a new analysis of these plates will provide valuable data for the dynamics of the Martian system. It is also important to notice that many useful observations are available in most of the observatories and national archives. Thus we have started to establish contacts for scientific purposes, and more particularly, to improve the dynamics of the planetary systems with old data, in the framework of the FP7 ESPaCE program and ahead of the upcoming NAROO project\footnote{See the NAROO webpage at \url{http://www.imcce.fr/hosted_sites/naroo/}} of the IMCCE. We also look forward to the arrival of the GAIA reference star catalog: reductions of old observations will yield increased accuracy by eliminating any errors caused by reference star positions.


\begin{acknowledgements}
     The research leading to these results has received funding from the European Community's Seventh Framework Program (FP7/2007-2013) under grant agreement n$\degr$263466 for the FP7-ESPaCE program.
\end{acknowledgements}

\bibliographystyle{aa}
\bibliography{mybiblio}

\end{document}